\documentclass[preprint2,natbib]{aastex}

\shorttitle{Optical and infrared photometry of the blazar PKS0537-441}
\shortauthors{Impiombato et al.}

\begin{document}

\title{Optical and infrared photometry of the blazar PKS0537-441:\\long and short time scale variability}

\author{D. Impiombato\altaffilmark{1,2,3}, S. Covino\altaffilmark{4,5}, A. Treves\altaffilmark{1,8}, L. Foschini\altaffilmark{4}, E. Pian\altaffilmark{6}, G. Tosti\altaffilmark{3},\\D. Fugazza\altaffilmark{4}, L. Nicastro\altaffilmark{7}, S. Ciprini\altaffilmark{3}}
\altaffiltext{1}{Universit\`a dell'Insubria, Dipartimento di Fisica e Matematica, Via Valleggio 11, 22100 Como, Italy}
\altaffiltext{2}{INAF, Osservatorio Astronomico di Padova, Via Vicolo dell'Osservatorio 5, 35122 Padova, Italy}
\altaffiltext{3}{Universit\`a di Perugia, Dipartimento di Fisica e Osservatorio Astronomico, Via A. Pascoli, 06123 Perugia, Italy}
\altaffiltext{4}{INAF, Osservatorio Astronomico di Brera, Via E. Bianchi 46, 23807 Merate (LC), Italy}
\altaffiltext{5}{INAF/TNG Fundaci\'on Galileo Galilei - Rambla Jos\'e Ana Fern\'andez P\'erez, 7, 38712, Bre\~{n}a Baja, TF - Spain}
\altaffiltext{6}{INAF, Osservatorio Astronomico di Trieste, Via G. B. Tiepolo 11, 34131 Trieste, Italy}
\altaffiltext{7}{INAF, Istituto di Astrofisica Spaziale e Fisica Cosmica, Via Gobetti 101, 40129 Bologna, Italy}
\altaffiltext{8}{Associated to INAF and INFN.}

\begin{abstract}
We present a large collection of photometric data on the Blazar PKS 0537-441
in the VRIJHK bands taken in 2004$-$2009. At least three flare-like episodes
with months duration, and $>$3 mag amplitude are apparent. The spectral
energy distribution is consistent with a power law, and no indication
of a thermal component is found. We searched for short time scale variability,
and an interesting event was identified in the $J$ band, with a duration of
$\sim$25 minutes.
\end{abstract}


\section{Introduction}
Blazars are active galactic nuclei (AGN) dominated by a relativistic jet pointing to the observer. The relativistic beaming enhances the continuum, and therefore the equivalent widths of spectral lines produced by photoionization of cold gas clouds around the nucleus may be modest, comparing with other AGN classes. In BL Lac objects, line weakness is extreme (equivalent width $<5$~\AA).

In order to relate dimensions and time scales in the jet comoving frame to what is directly observed, one should go through Lorentz transformations, which are characterized by the Doppler factor defined as $\delta=\Gamma^{-1}(1-\beta \cos \theta)^{-1}$, $\Gamma$ is the Lorentz factor and $\theta$ is the angle of sight. When the relevant dimensions are much smaller than the spatial angular resolution, which is the usual case for the inner jet, the properties of the jet can be effectively constrained by the study of the variability of the emission \citep [e.g.][]{ulrich97}.\\

Blazars are generally observed in a large frequency range, which, in some cases, covers the entire electromagnetic spectrum from the radio band to the TeV gamma rays (e.g. Fossati et al. 1998, Donato et al. 2001, Ghisellini et al. 2010). The basic model for explaining the spectrum of blazars is in terms of synchrotron radiation, plus an inverse-Compton (IC) component, which may be due to scattering off the electrons of the synchrotron photons, or those coming from the accretion disk, broad-line region or infrared torus. In practice, spectra are well constrained if the source size can be obtained though variability studies. The ideal situation would be to obtain the time dependence of the spectral energy distribution in the entire band of observation, and this has been done for very few objects (e.g. Hartmann et al. 2001, Ballo et al. 2002, Acciari et al. 2009, Fuhrmann et al. 2010).

The  systematic study of blazar variability in the optical and infrared bands has greatly profited of the development of robotic telescopes, which ensure long observing campaigns, with relatively little intervention from the observer. The large observing time can yield information complementing that of multifrequency campaigns, which obviously can cover at most few days.

Here we report on 2004$-$2009 six filters photometry of the blazar PKS 0537-441, which is unprecedented for the observing time dedicated to the source. PKS 0537-441 is a $z=0.896$ bright (V$\sim$13) blazar, which was studied from radio to GeV $\gamma$ rays. The main characteristics of the broad-band spectrum are the synchrotron peak in the IR, the IC hump with a maximum in the $\gamma$ rays, the X-ray spectrum sampling the tail of the synchrotron and the rise of the IC, and the presence of emission lines in the optical spectrum. Therefore, the object belongs to the low-frequency peaked subclass of blazars (Pian et al. 2002, 2007). 

In Section 2, we describe the observational procedures, data reduction and analysis. The results are given distinguishing between long- and short-term variability (Sections 3 and 4). In the discussion section in particular we focus on the relevance of short time scales of variability in constraining the beaming of the source.

\section{Telescope and data analysis}
The Rapid Eye Mount (REM) telescope is a 60\,cm fully robotic instrument located at the ESO Cerro La Silla observatory (Chile). The telescope has a Ritchey-Chretien configuration with a 60 cm f /2.2 primary and an overall f /8 focal ratio in a fast moving alt-azimuth mount providing two stable Nasmyth focal stations. At one of the two foci, the telescope simultaneously feeds, by means of a dichroic, two cameras: REMIR for the Near-InfraRed (NIR) \citep[see][]{conconi04} and ROSS \citep[see][]{tosti04} for the optical. Both the cameras have a field of view of $10^{\prime}\times10^{\prime}$ and imaging capabilities with the usual NIR ($z^\prime$, J, H, and K) and Johnson-Cousins VRI filters. The telescope was originally designed for follow-up of \textit{Swift} gamma-ray bursts, but for most of the time it is used for standard observational programs. More information about the REM project and capabilities can be found in Zerbi et al. (2001), Chincarini et al. (2003), Covino et al. (2004).

Both REMIR and ROSS instruments were used in order to obtain nearly simultaneous data and to study the spectral behavior of PKS0537-441 at different levels of flux. The REM telescope has systematically observed this source for 5 years between December 2004 and March 2009.

Data analysis has been carried out following standard recipes. Instrumental magnitudes were obtained via aperture photometry by means of an automatic pipeline\footnote{http://www.merate.mi.astro.it/utenti/covino/usermanual.html} which aligns and rotates frames to a common reference, derives and calibrates photometry comparing to standard stars in the field from the 2MASS catalogue\footnote{http://irsa.ipac.caltech.edu} in the NIR and from sequences reported by \citet{hamuy89} in the optical. Results were controlled manually using the Graphical Astronomy and Image Analysis (GAIA)\footnote{http://star-www.dur.ac.uk/$\sim$pdraper/gaia/gaia.html} tool: we selected only frames with reliable photometry and absolute calibration based on the presence of the calibration stars, on their consistency, stable background, correct centering, etc. In particular, for all frames, we required that at least three out four calibration stars were detected. The reference stars, the same in the optical and in the NIR, are stars Nr. 1, 2, 3 and 6 as labeled in Hamuy \& Maza (1989) corresponding to the 2MASS stars $05390451-4406382$, $05384933-4401300$, $05385997-4401191$ and $05383780-4403241$, respectively.  The agreement between the automatic pipeline and the manual analysis is typically better than 10\% in flux. 

In the following, fluxes are corrected with $A_V=0.18$ (Kalberla et al. 2005) and the extinction laws of Cardelli et al. (1989), while magnitudes are the observed ones. The quoted errors are at 1$\sigma$ and they include photometric and calibration uncertainties.

\section{Long time scale variability}

\begin{figure*}
\centering
\begin{tabular}{cc}
\includegraphics[angle=270,width=7cm]{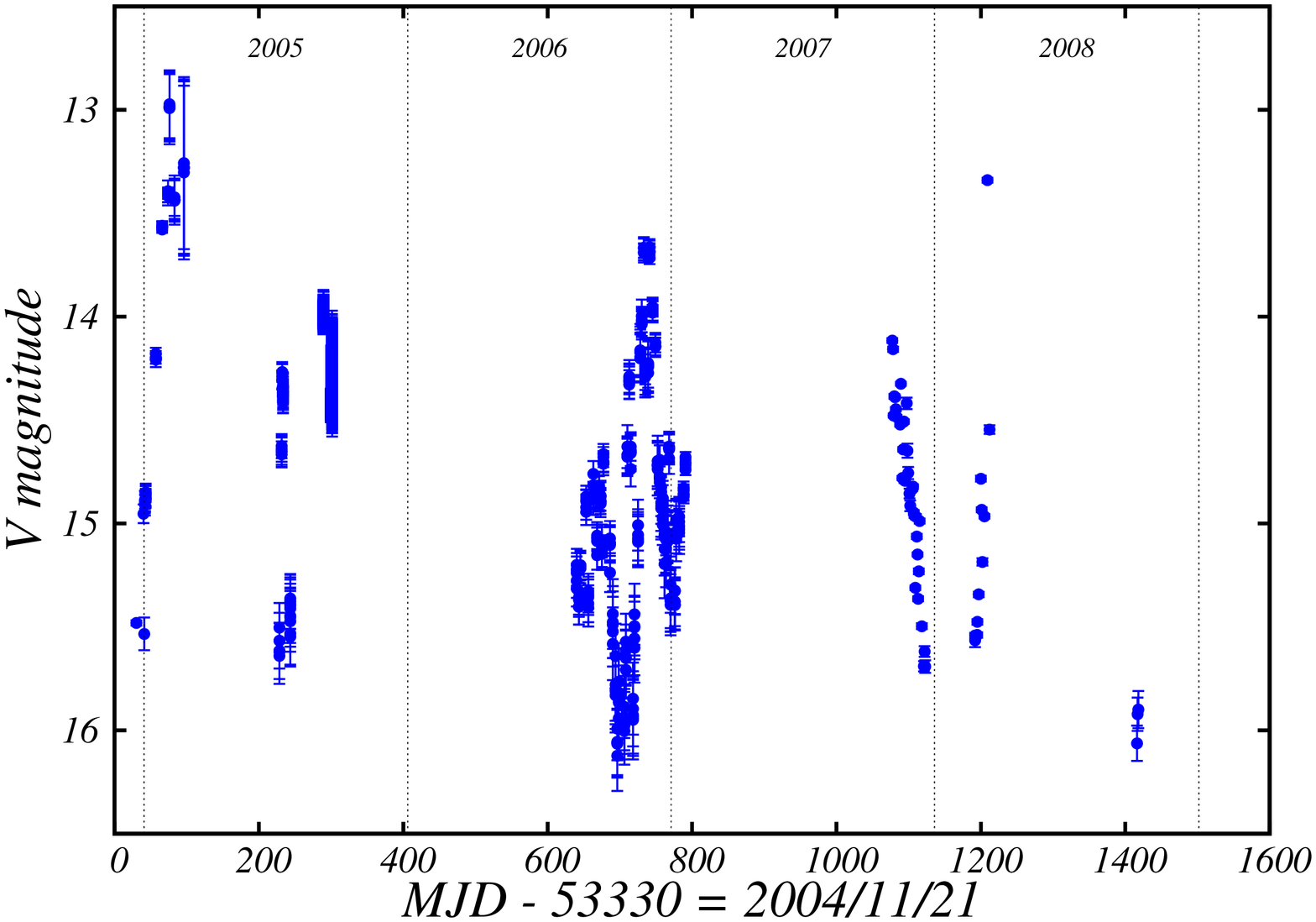} &
\includegraphics[angle=270,width=7cm]{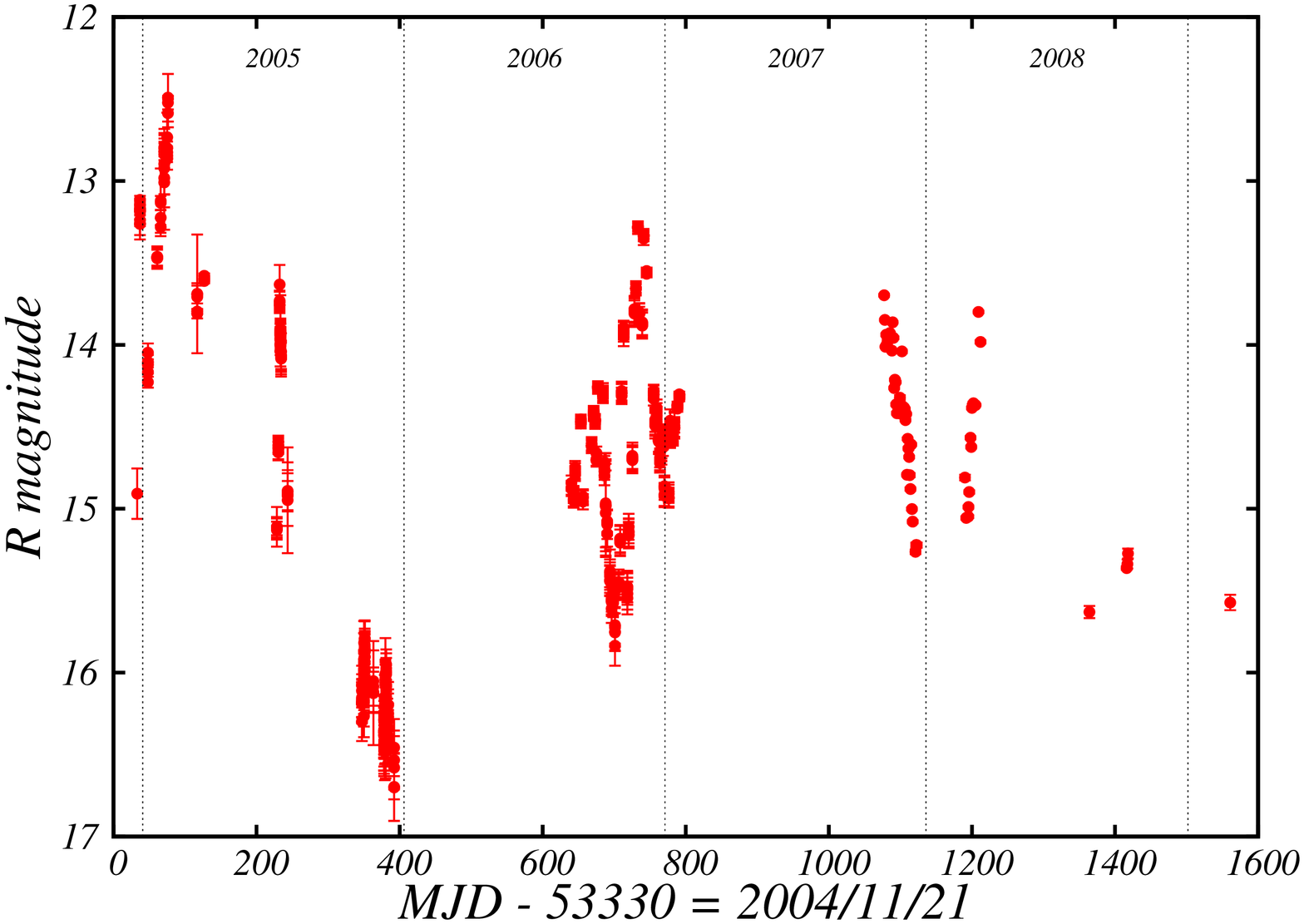} \\
\includegraphics[angle=270,width=7cm]{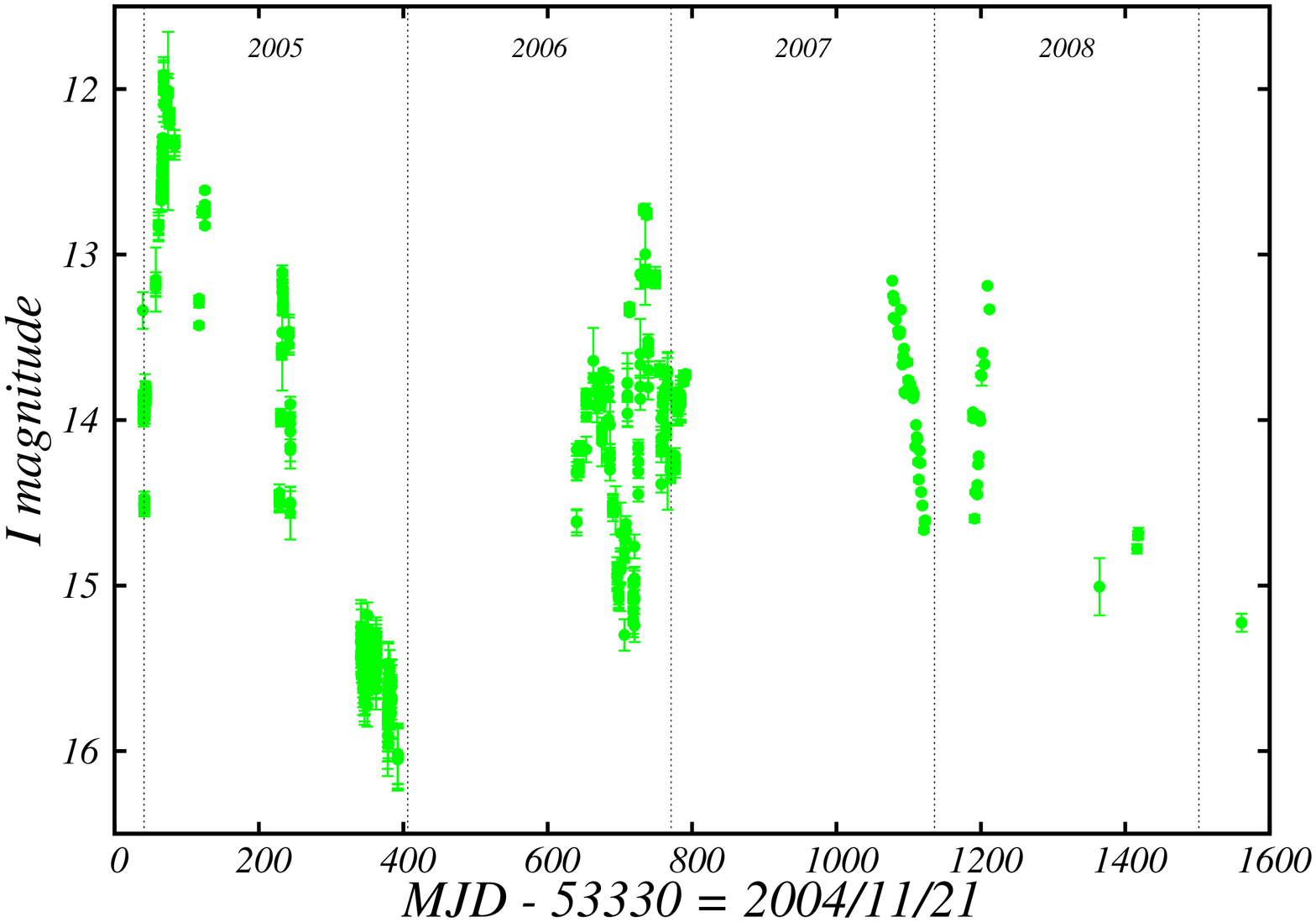} &
\includegraphics[angle=270,width=7cm]{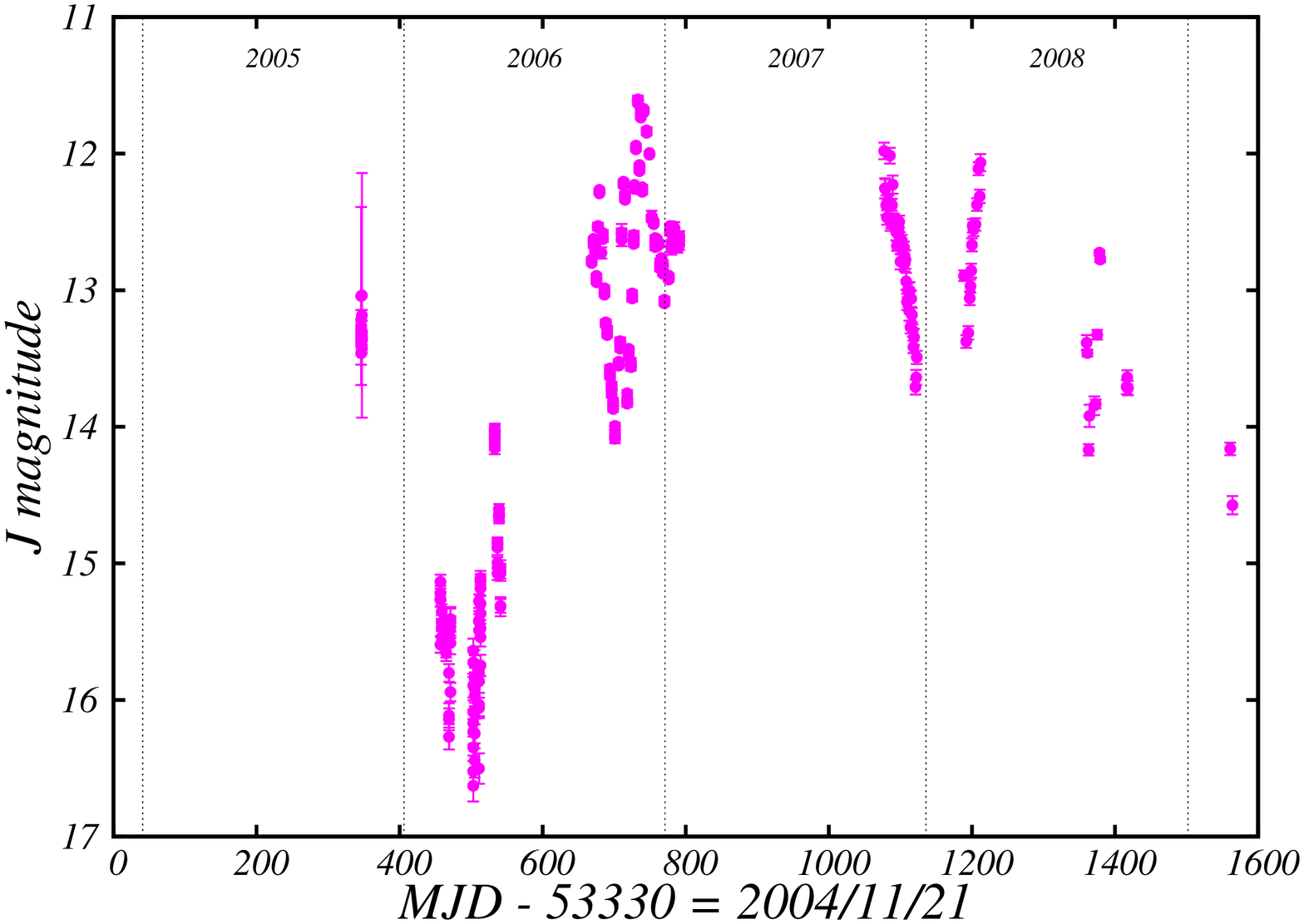} \\
\includegraphics[angle=270,width=7cm]{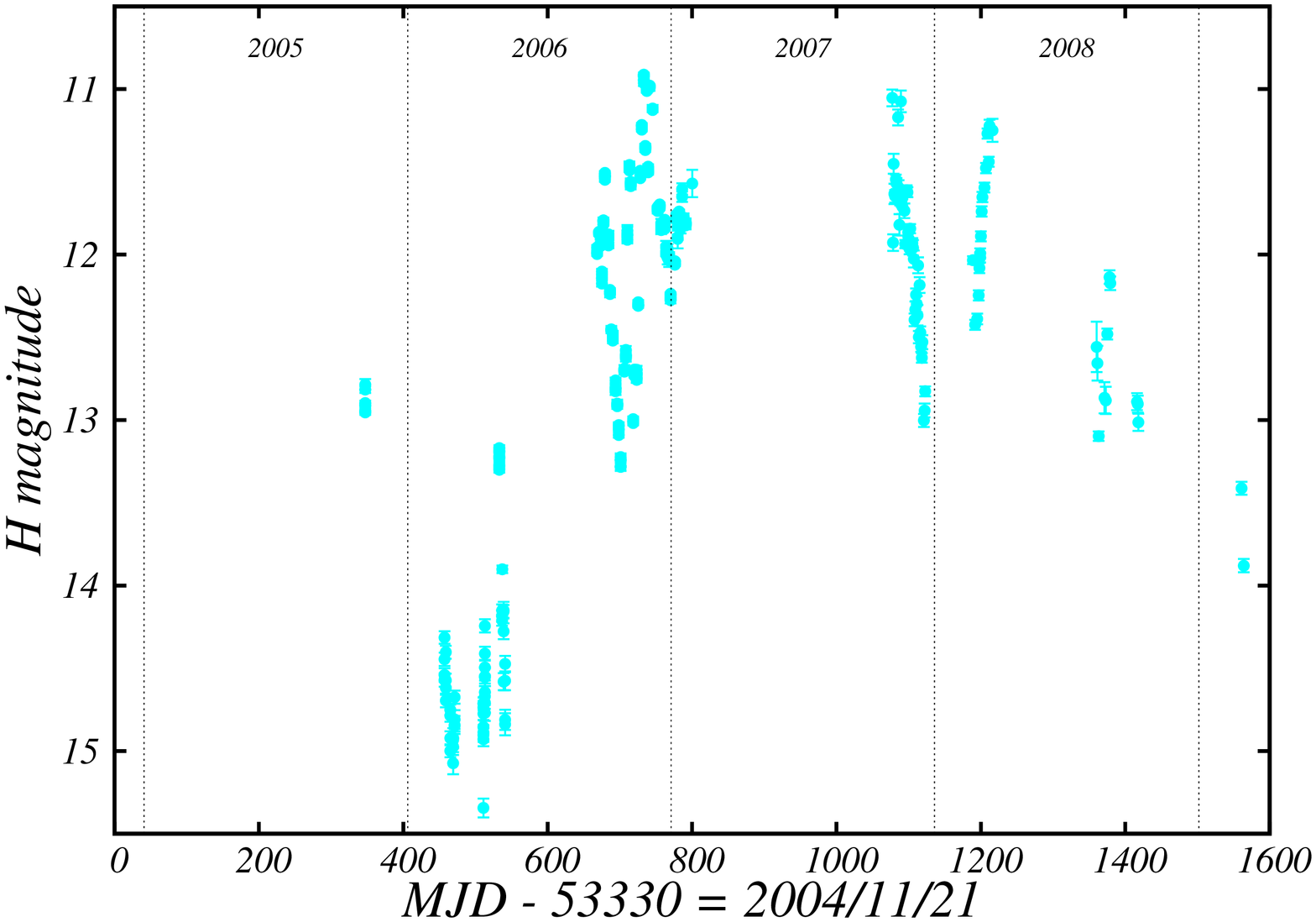} &
\includegraphics[angle=270,width=7cm]{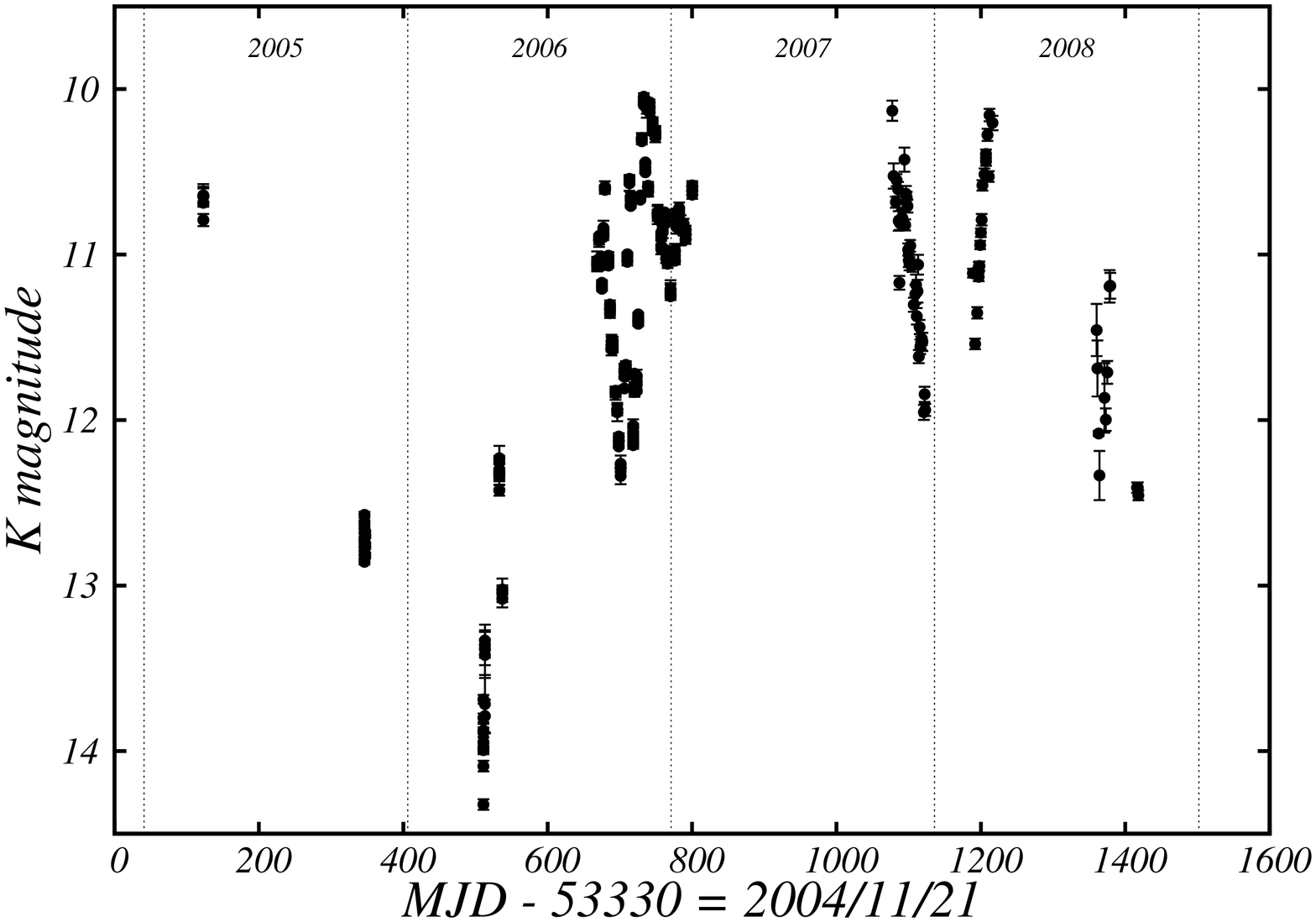} \\
\end{tabular} 
\caption{Optical and NIR light curves of PKS0537-441 in the 2004$-$2009.}
\label{fig:opticalNIR}
\end{figure*}

\begin{figure*}
\centering
\begin{tabular}{cc}
\includegraphics[angle=270,width=7cm]{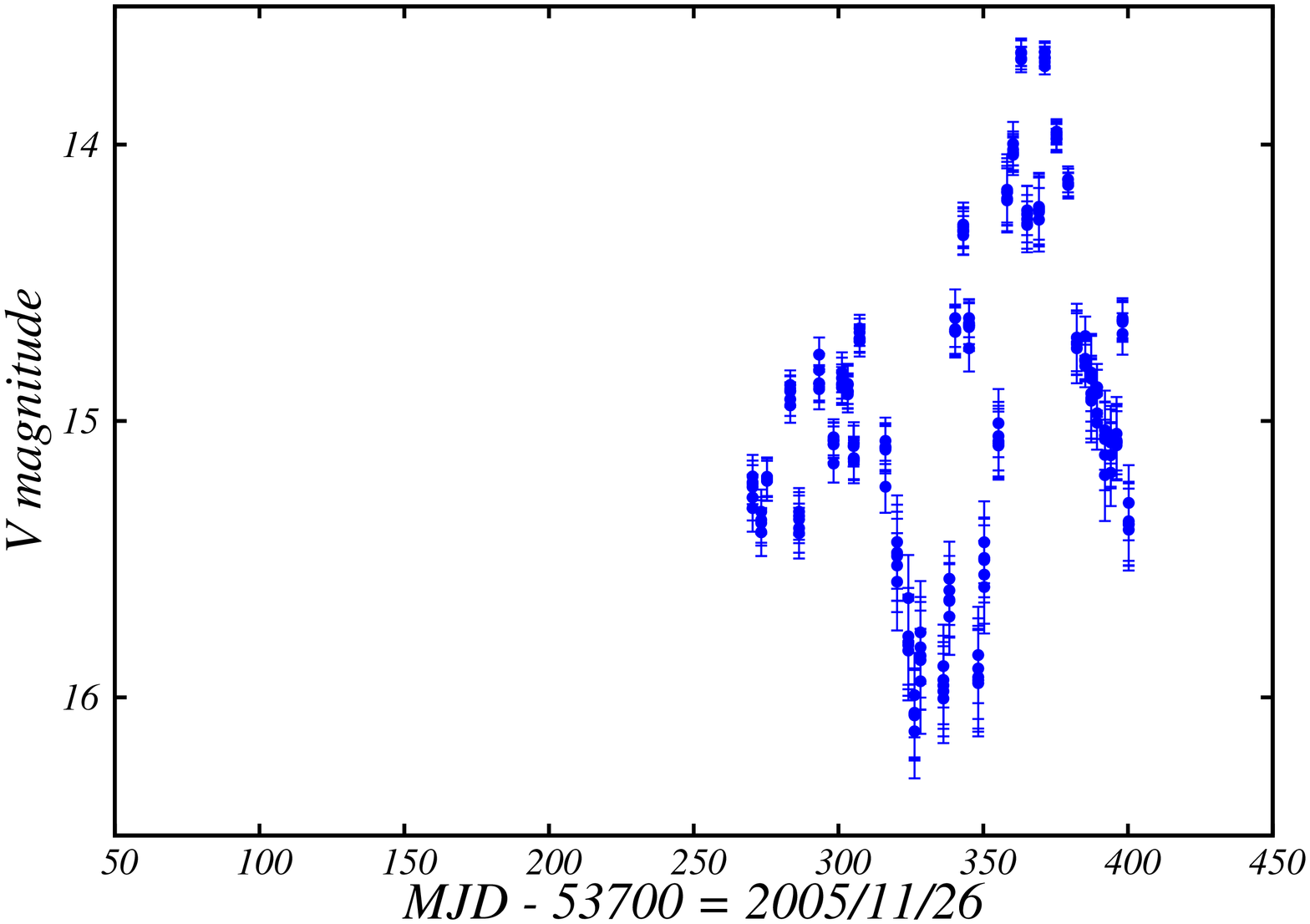} &
\includegraphics[angle=270,width=7cm]{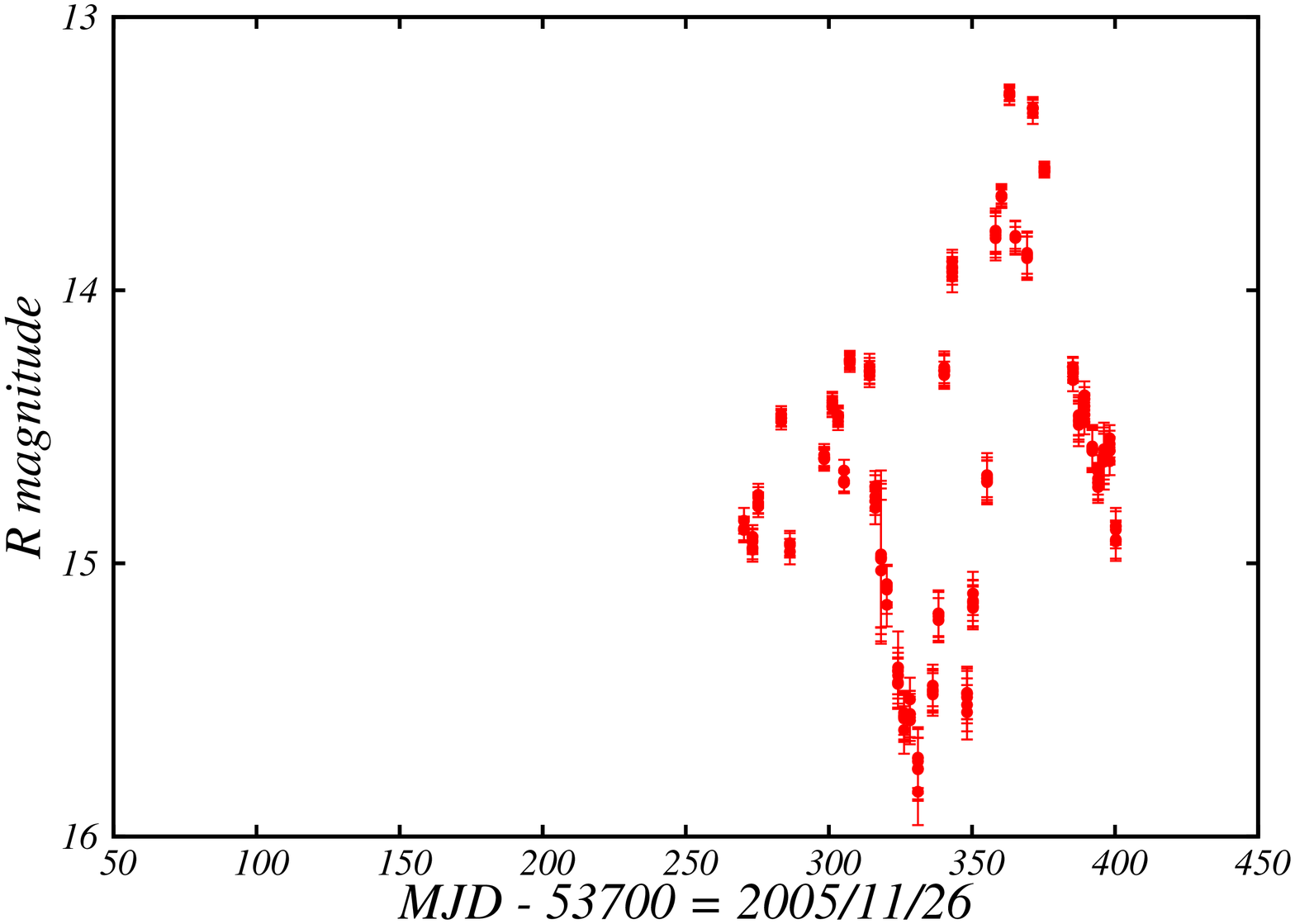} \\
\includegraphics[angle=270,width=7cm]{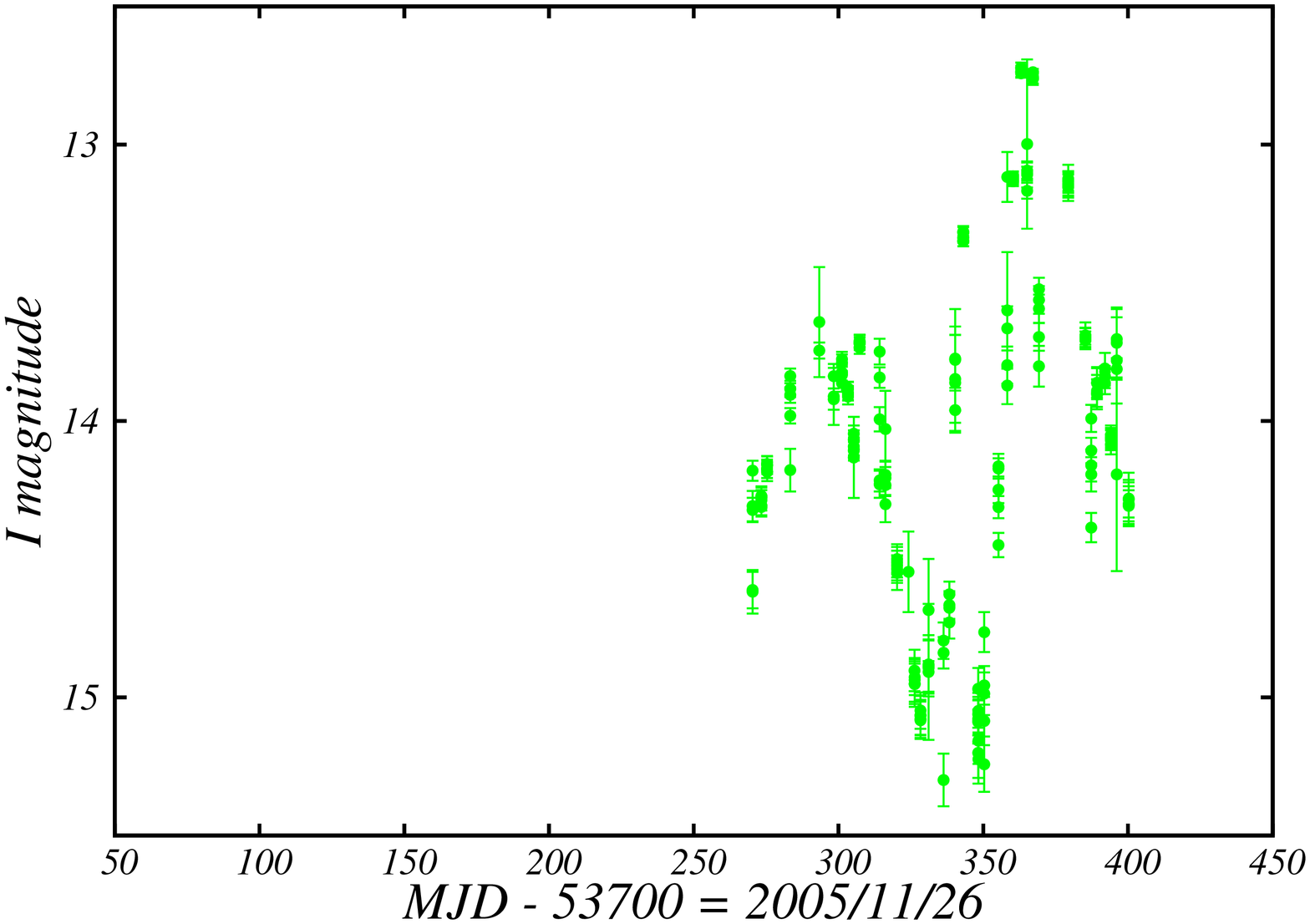} &
\includegraphics[angle=270,width=7cm]{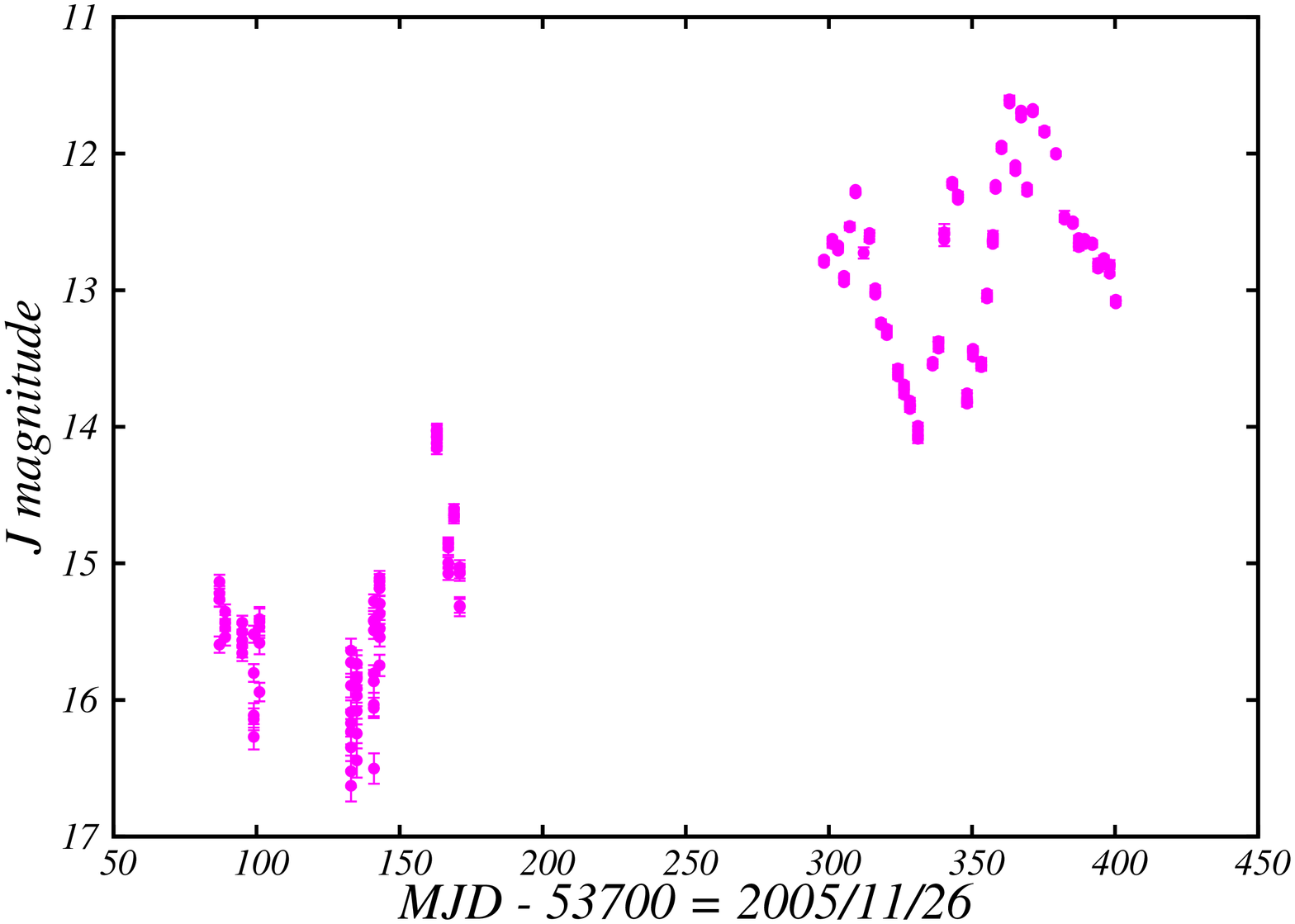} \\
\includegraphics[angle=270,width=7cm]{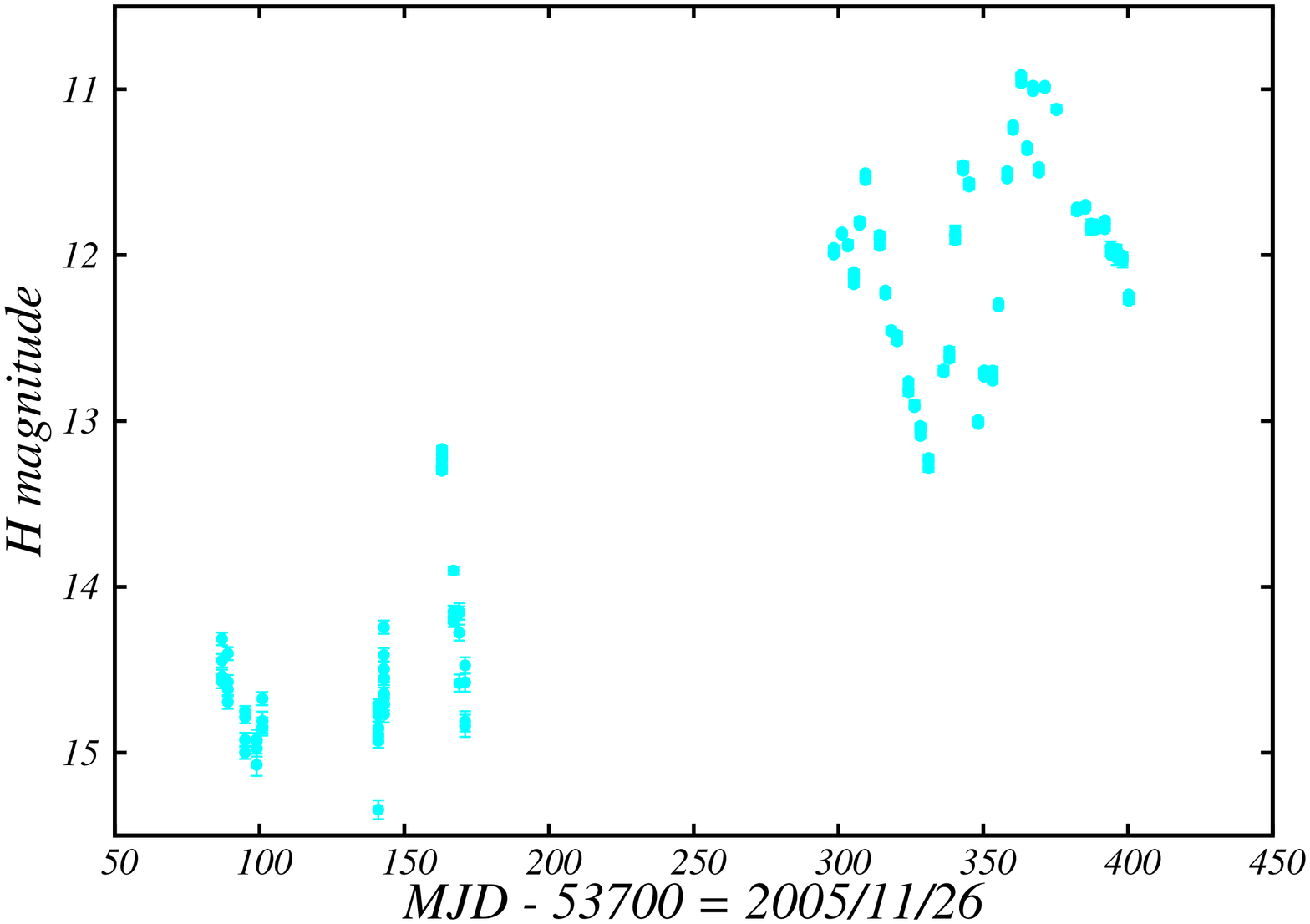} &
\includegraphics[angle=270,width=7cm]{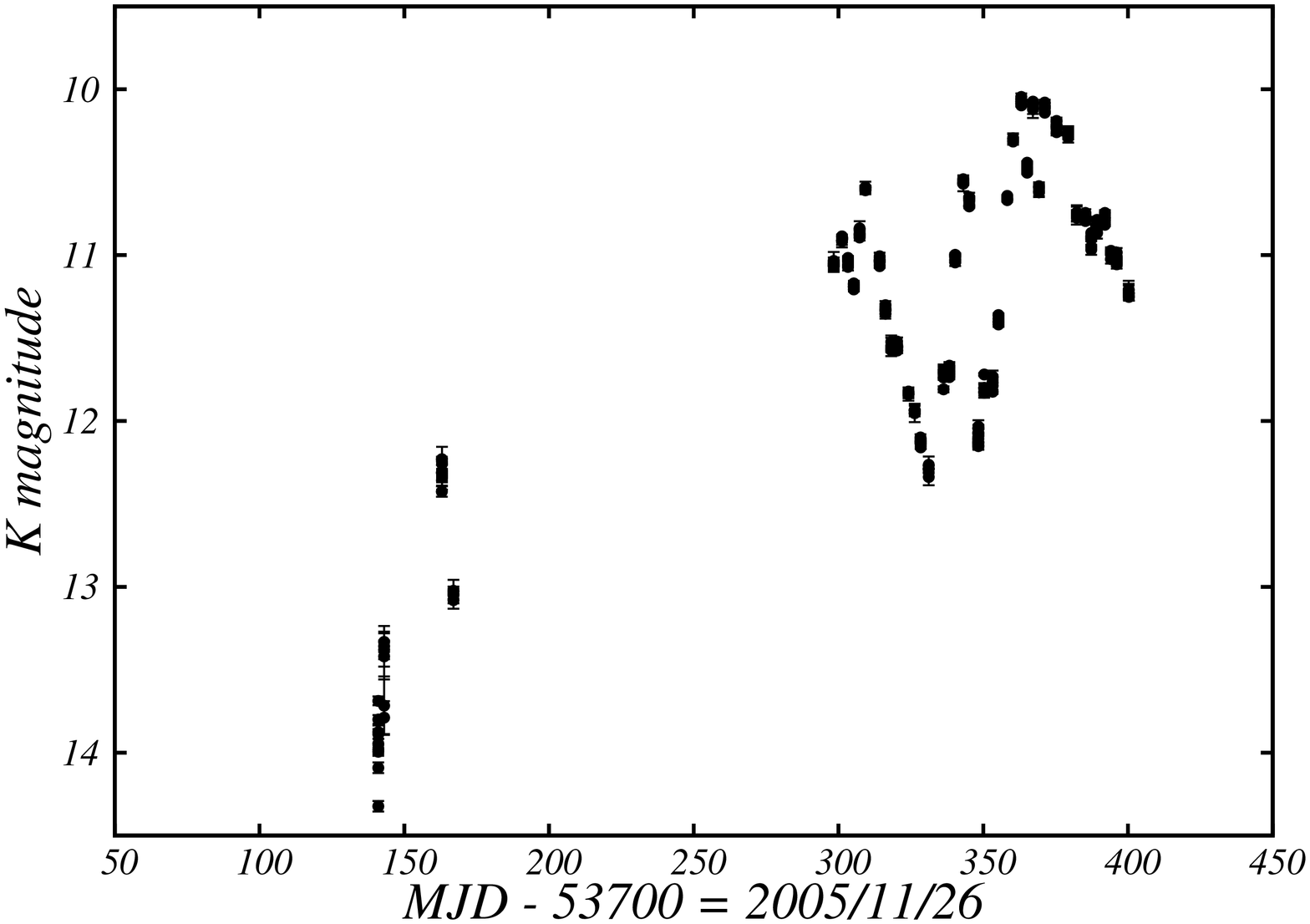} \\
\end{tabular}
\caption{Optical and NIR light curves of PKS0537-441 in the year 2006.}
\label{fig:zoomONIR}
\end{figure*}

Fig. \ref{fig:opticalNIR} represents the light curves of the source in the six bands. Part of these data was already published in Dolcini et al. (2005), Pian et al. (2007), Impiombato et al. (2008). Pucella et al. (2010) consider some photometry of the source in October 2008 coincident with an \emph{AGILE} $\gamma-$ray observation. The 2006 data, which are noticeable for the large and complex variability, are reported in an enlarged version in Fig. \ref{fig:zoomONIR}. For all photometric points (see Table \ref{tab:examplephot}), we give the filter, the reference MJD at the beginning of the exposure, the integration time, the magnitude, and the corresponding error. The typical integration times are 20$-$600 s in the optical and 170 s in infrared bands.

The maximum and minimum states are reported for each filter in Table \ref{tab:maxmin}. The values have been controlled through the manual analysis and are in good agreement with the automatic analysis ones. The variability ranges from $\sim 4$~mag in V to $\sim 6$~mag in J. Note that the time coverages in the various filters are not equal.

\begin{table*}[h!]
\centering
\caption{Example of photometry of PKS0537-441 from December 2004 to March 2009. The full version of the table is available online.\label{tab:examplephot}}
\smallskip
\begin{tabular}{c c c c c }
\hline\hline
MJD & Exposure & Filter & Magnitude & Error\\
\hline
d & s & & mag & mag  \\
\hline

53357.17192	&	600	&	V	&	16.20	&	0.08	\\
55326.98610	&	300	&	V	&	14.02	&	0.11	\\
... & ... & ... & ... & ... \\
\hline
53360.16350	&	30	&	R	&	14.93	&	0.15	\\
55327.01266	&	300	&	R	&	13.88	&	0.03	\\
... & ... & ... & ... & ... \\
\hline
53358.18466	&	600	&	I	&	15.28	&	0.02	\\
55327.01645	&	300	&	I	&	13.43	&	0.06	\\
... & ... & ... & ... & ... \\
\hline
53676.17050	&	30	&	J	&	13.04	&	0.05	\\
55326.99834	&	30	&	J	&	12.37	&	0.03	\\
... & ... & ... & ... & ... \\
\hline
53677.29373	&	30	&	H	&	12.62	&	0.26	\\
55327.01354	&	30	&	H	&	11.61    &	0.05	\\
... & ... & ... & ... & ... \\
\hline
53453.02882	&	30	&	K	&	10.71	&	0.04	\\
55312.04763	&	30	&	K	&	10.59	&	0.05	\\
... & ... & ... & ... & ... \\
\hline
\end{tabular}
\end{table*}

\begin{table*}[ht!]

\caption{Maximum and minimum intensity state of PKS0537-441.\label{tab:maxmin}}
\centering
\begin{tabular}{c c c c c }
\hline\hline
MJD & filter & Magnitude & Flux(mJy)&$\Delta$Magnitude \\
\hline
55255.22365  &  V  & 12.95$\pm$0.09 & $24.1 \pm 2.0$  \\
53677.29965  &  V  & 16.87$\pm$0.25 & $0.65 \pm 0.15$  \\
             &  V  &                &               & 3.919$\pm$0.266 \\
\hline
55292.11168  &  R  & 12.08$\pm$0.07 & $42.1 \pm 2.6 $  \\
53726.23140  &  R  & 16.80$\pm$0.22 & $0.55 \pm 0.11$  \\
             &  R  &                &               &4.716 $\pm$0.233\\
 \hline
53406.11486  &  I  & 12.00$\pm$0.08 & $35.8 \pm 2.5$  \\
53726.35053  &  I  & 16.37$\pm$0.24 & $0.64 \pm 0.14$   \\
             &  I  &                &               &4.37 $\pm$0.254\\
\hline
55271.07350  &  J  & 10.78$\pm$0.12 & $77.5 \pm 8.9$  \\
53834.98987  &  J  & 17.78$\pm$0.25 & $0.12 \pm 0.03$  \\
             &  J  &                &               &6.998 $\pm$0.274\\
\hline
54063.13561  &  H  & 10.86$\pm$0.03 & $46.3 \pm 1.4$  \\
53835.04113  &	H  & 15.99$\pm$0.05 & $0.41 \pm 0.02$  \\
             &  H  &                &               &5.129 $\pm$0.062 \\
\hline
54063.14019  &  K  & 10.04$\pm$0.03 & $64.5 \pm 1.6$  \\
53835.01757  &	K  & 15.37$\pm$0.06  & $0.47 \pm 0.03$  \\
             &  K  &                &               &5.334 $\pm$0.068\\
 \hline
 \end{tabular}
 \end{table*}

\section{Short time scale variability}
Since the typical exposure duration is 20$-$600 seconds, variability on a few minute time scale can be searched for. However, our observing campaigns were not optimized for the study of such variability and, in fact, there are relatively few epochs where 5$-$10 points within an hour are available. Moreover, the observing procedure favours filter shifts after few photometric points. Nonetheless, we have considered the usual definition of an exponential increase/decrease of flux: 

\begin{equation}
F(t)= F(t_0)\exp[-(t-t_0)/\tau]
\label{eq:timescale}
\end{equation} 

where $F(t)$ and $F(t_0)$ are the fluxes at the time $t$ and $t_0$, respectively, and $\tau$ is the characteristic time scale. The time scale $\tau$ intrinsic to the source is then calculated as $\tau_{\rm int}=\tau/(1+z)$.

We considered only frames with reliable photometry basing on consistency among the various reference stars (at least three) and no image defects close to the target position.  An automatic procedure was launched in order to find events of minimal time scale, $\tau$ ($< 1$~hour), with flux differences with a significance above $5\sigma$. One event was recovered, which appear of interest. It is reported in Table~\ref{tab:shortevents}: all the frames were carefully checked manually. No simultaneous events satisfying our criteria were detected in the optical bands. Note that the telescope is designed to provide the best sensitivity in the NIR.

\setcounter{table}{2}
\begin{table*}[th!]
\centering
\caption{A short event. \label{tab:shortevents}}
\begin{tabular}{c c c c c c}
\hline\hline
MJD & Filter & Flux & $\Delta$T & $\tau_{\rm int}$ & Significance\\
\hline
    &        & mJy  & min         & min  & $\sigma$\\
 \hline

\hline
54038.29736 & J  & 7.22$\pm$0.09 & & &\\
54038.29944 & J  & 7.69$\pm$0.06 & & &\\
              & J  &               & 3  & 25.2$\pm$0.3 & 5.4\\
\hline
 \end{tabular}
 \end{table*}

\begin{figure*}
\centering
\includegraphics[angle=-90,scale=0.6]{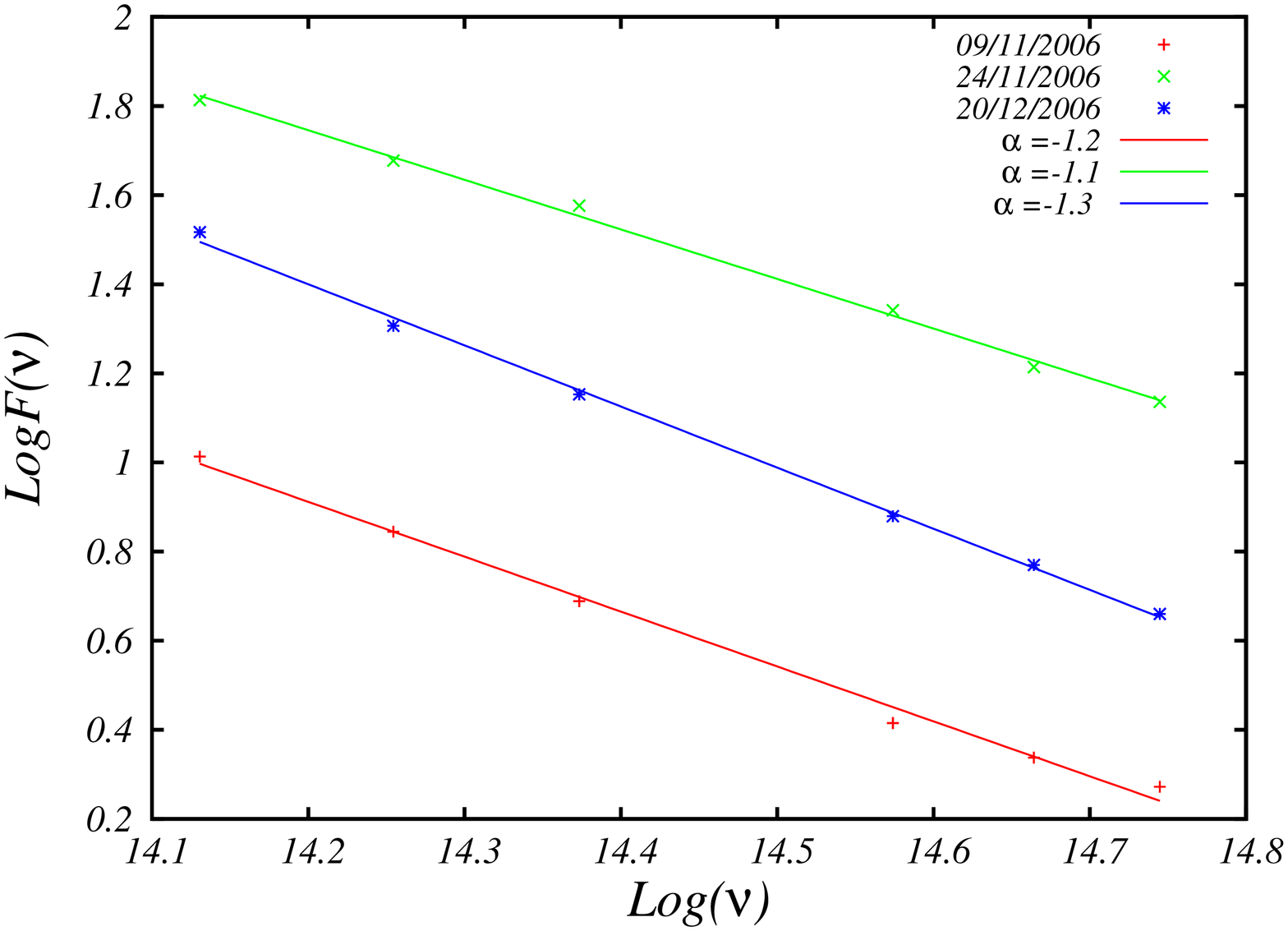}
\caption{Spectral flux distribution at three epochs. Fluxes $F_{\nu}$ are in mJy, frequency $\nu$ in Hz.}
\label{fig:SED}
\end{figure*}

\section{Discussion}
The light curves of 2004$-$2009 (Figs. \ref{fig:opticalNIR} and \ref{fig:zoomONIR}) correspond to rather irregular exposures and are interrupted by large gaps, partly due to the visibility conditions of the source. Nonetheless, this is the largest collection of published data on PKS 0537-441 in terms of overall exposure ($t \sim 3 \times10^{5}$ s) and number of photometric points in all bands (about 4000 points). 

A visual inspection of Figs. \ref{fig:opticalNIR} and \ref{fig:zoomONIR} indicates very structured light curves, with grossly similar shapes in all bands. A number of rapid brightenings and fadings, which may suggest flare like episodes, are apparent. Typical durations are months and the amplitudes are of several magnitudes (see also Table~\ref{tab:maxmin}). The event described by \citet{dolcini05} and \citet{pian07} is that of Jan 2005 (days 41$-$71 in VRI bands, see Fig. \ref{fig:opticalNIR}). That of February$-$March 2008 (days 1189$-$1216 in VRIJHK bands, see Fig. \ref{fig:opticalNIR}) is the one considered by \citet{impiombato08}. A double flare structure is present in the period 21/02/2006$-$31/12/2006 (see Fig. \ref{fig:zoomONIR}). Probably the event of 1971, reported by \citet{eggen73}, 4 mag in V in 8 months, is analogous to those examined here. 

We compared the spectral shapes at three epochs where all six filters are available and the intensity states are significantly different. An interesting case is shown in Fig. \ref{fig:SED}, where it is apparent that the SED can be satisfactorily described by power laws. It is noticeable that the higher and lower states, which are separated by two weeks, exhibit an intensity variation by a factor $\sim 10$ and the spectral index changed only slightly ($\Delta \alpha \sim 0.1$). Therefore, in both the low and high state the emission appears dominated by the jet, and no evidence of a thermal component is apparent. Some deviations from a power law may appear in the low-state 2004 December 21 \citep[Fig. 3 of][]{dolcini05}. The thermal component mentioned by \citet{pian07} is probably a consequence of the extreme variability of the source.

We now concentrate on the short time scale variability, and in particular on the event of 2006 October 30 (MJD~54038). This is rather rapid: comparable to what observed by Villata et al. (2008) in the notorious highly variable BL\,Lac object S5~0716+714, about 0.002 mag/min in V. A time scale $\tau_{\rm int}\sim 25$~min corresponds to a size in the source frame  $R'< c \tau_{\rm int} \sim 4 \times 10^{14}$~cm, having considered $\delta \sim 10$, typical of blazars (e.g.  Ghisellini et al 2010, and references therein). This can be compared with other constraints on the size of the source. Indeed, as shown in Table 4 of \citet{pian05}, the mass estimates of PKS 0537-441 range from $0.5 \times 10^{8}M_{\odot}$ to $16\times 10{^8}M_{\odot}$. They are based on the measurements of line widths and the continuum intensity, which enable one to deduce the velocity of the clouds emitting the broad lines and their distance. The mass then follows from the virial theorem. The problem is that the continuum is dominated by the jet and one should use relativistically corrected values. Such a correction is introduced, but in a somewhat gross formulation, only by \citet{pian05}, who favors the lowest value of mass. The gravitational radius $R_g=GM/c^2$ for masses in the $10^{8-9}M_{\odot}$ is $1.5\times 10 ^{13-14}$ cm. Taking a jet basis at 6 gravitational radii one has typical dimensions in the range $2\times 10^{14-15}$ cm. These results confirm that the blazar duty cycle is able of extremely rapid changes in flux both at $\gamma$ rays and optical frequencies and can have strong impact in the evaluation of the cosmic high-energy background (Giommi et al. 2006, Draper \& Ballantyne 2009).

Variability on time scale of minutes in the optical band appears a rather accessible probe of the relativistic jet structure. In future observations of PKS 0537-441 and similar sources one should take into account that rapid variability of the type reported here appears a rare event and therefore very long monitoring should be  scheduled. An important improvement with respect to our photometry would be the actual simultaneity in all bands, which can be obtained by beam splitting, since one could probe the color of the short events.

\end{document}